\documentclass[12pt]{iopart}
\usepackage{iopams}
\usepackage{hyperref}
\usepackage{graphicx}
\usepackage{cite}
\newcommand{\D}{{\rm d}}
\newcommand{\ds}{\displaystyle}
\newcommand{\bea}{\begin{eqnarray}}
\newcommand{\eea}{\end{eqnarray}}
\usepackage{color}

\usepackage{epsfig}

\begin{document}

\title[An exactly solvable predator prey model with resetting]
{An exactly solvable predator prey model with resetting}

\author{Martin R. Evans$^1$, Satya N. Majumdar$^2$ and Gr{\'e}gory Schehr$^3$}

\address{$^1$ SUPA, School of Physics and Astronomy, University of Edinburgh, 
Peter Guthrie Tait Road, Edinburgh EH9 3FD, UK\\
$^2$ LPTMS, CNRS, Univ. Paris-Sud, Universit\'e Paris-Saclay, 91405 Orsay, France \\
$^3$ Sorbonne Universit\'e, Laboratoire de Physique Th\'eorique et Hautes Energies, CNRS UMR 7589, 4 Place Jussieu, 75252 Paris Cedex 05, France
}
\ead{m.evans@ed.ac.uk,satya.majumdar@u-psud.fr,gregory.schehr@u-psud.fr}


\date{\today}
\begin{abstract}
We study a simple model of a diffusing particle (the prey)
that on encounter with one of a swarm of diffusing  predators can either perish or be reset to its original position at the origin. We show that the survival probability of the prey up to time $t$ decays algebraically as $\sim t^{-\theta(p, \gamma)}$ where the exponent $\theta$ depends continuously on two parameters of the model, with $p$ denoting the probability that a prey survives upon encounter with a predator and $\gamma = D_A/(D_A+D_B)$ where $D_A$ and $D_B$ are the diffusion constants of the prey and the predator respectively. We also compute exactly the probability 
distribution $P(N|t_c)$ of the total number of encounters till the capture time $t_c$ and show that it exhibits an anomalous large deviation form $P(N|t_c)\sim t_c^{- \Phi\left(\frac{N}{\ln t_c}=z\right)}$ for large $t_c$. The rate function $\Phi(z)$ is computed explicitly. Numerical simulations are in excellent agreement with our analytical results.
\end{abstract}

\section{Introduction}

It is by now well established that resetting a stochastic process to its initial condition can fundamentally change the  behaviour of the process. For example, the introduction of  resetting can generate nontrivial nonequilibrium stationary states and can
strongly affect first passage properties \cite{EMS20}.

An archetypal example of a stochastic process under resetting is a diffusing particle that is reset to the origin  after random waiting times with exponential distribution (Poissonian resetting) \cite{EM1,EM2,EM14,MSS15a}. It has been shown that the mean first-passage time to some target
is rendered finite, rather than infinite as in the absence of resetting.
Moreover, under resetting the survival probability of the diffusive particle in the presence of  an absorbing target decays  exponentially  in time,  whereas without resetting
the decay is a well-known power law $\sim t^{-1/2}$.
Various aspects and generalisations of diffusion under resetting have been studied both theoretically (see for example \cite{MV13,WEM13,BS14,Reuveni14,GMS14,KMSS14,MSS15,MST15,EM16,PKE16,NG16,Reuveni16,BEM17,PR17,RG17,HT17,MO18,MSM18,CS18,EM19,BKP19,HMMT19,MPCM19,MM19,Gr20,Bressloff20,MMS20,MMSS21,CE21}) and experimentally in optical traps~\cite{Tal20,Besga20,Faisant21}.

Our aim in this work is to illustrate how  resetting can also create non-trivial power-law decays for a survival probability, with exponents that vary continuously on the parameters of the model. In the case of diffusion with resetting a power-law decay occurs for the average survival probability of a static target at the origin in the presence of a finite density of  diffusive particles \cite{EM1}.
The diffusive particles each  reset to their own initial positions. Interestingly,  the typical survival probability decays exponentially and  the power-law decay of the average is due to rare configurations of the initial positions  \cite{EM1}. It is of interest to explore other  mechanisms to generate power-law decays.

In this work we consider a simple model of a diffusing particle (the prey)
that on encounter with one of a swarm of diffusing  predators can either perish or be reset to its original position  at the origin. Without resetting, the first-passage properties of a swarm of particles was studied in Ref. \cite{MOS11}. Here, we consider the additional effect on this swarm of particles (predators) when the target (prey) has a resetting dynamics.  
Furthermore,  the reset is triggered by some internal dynamical event  rather than an external process, as in Poissonian resetting.
Related  scenarios that have previously been considered are
two diffusing particles, which reset to their initial positions on contact \cite{FE17},
and  first-passage resetting where a reset event occurs when the diffusing particle reaches a specified threshold \cite{DBRFR20}.

As we shall see,  our predator prey model effectively reduces to a diffusive process that on reaching the origin has a 
probability of resetting to a time-dependent, Gaussian distribution. The resets slow down the  absorption of the process and
turns the temporal decay of the survival probability into a power-law form with a non-trivial exponent
that depends continuously on the parameters of the model.
This  leads us naturally to investigate the statistics of the number of resetting events up until eventual capture.
We compute the joint distribution of the number of resets $N$ and capture time $t_c$.
It turns out that this has an interesting scaling
with the mean and variance of the number of resetting events depending logarithmically on the capture time. Moreover, we are able to compute the large deviation function for  the probability  of the number of resets, conditioned on the capture time.

The paper is organised as follows. In Section 2 we define the predator-prey model and
compute the Laplace transforms of the survival probability of the prey and the capture time distribution. From these we extract the asymptotic  power-law decay.
In Section 3  we consider the statistics of the number of encounters until the time that capture occurs  and compute the Laplace transform of the probability of $N$ encounters given the capture time $t_c$. From this we extract the asymptotics of the
moments of $N$. In Section 4 we compute the large deviation form of $P(N|t_c)$ and show that it translates into an asymptotic  power-law decay for large $t_c$
\begin{equation}
P(N|t_c)\sim t_c^{- \Phi\left(\frac{N}{\ln t_c}=z\right)}\, ,
\label{ldv_tc.0}
\end{equation}
where the large deviation function $\Phi(z)$ acts as an effective exponent.

\section{A simple predator-prey model with resetting}

We consider a single prey and a swarm of predators diffusing on a line with diffusion constants $D_A$ (for the prey)
and $D_B$ (for each of the predators). We assume that the population of the predators in the swarm is infinite and we label them $n=1, 2, 3, \ldots$. Initially, only the predator number $n=1$ is ``active'' in the sense that only this predator can detect the prey. The other predators are ``passive'', i.e., even if
they cross the path of the prey, they do not detect it. When the prey and the current active predator encounter each other,
with probability $p$ the active predator retires and the prey escapes to its retreat or safe haven located at the origin. 
With the complementary probability $1-p$, the predator
wins, in which case the prey dies. In the former case, when the prey escapes, another predator (say $n=2$) becomes the current
active predator and the pursuit continues. We ask the simple question:  what is the probability of the prey surviving up to some time $t$?

\subsection{Model definition}
More precisely, let $x_A(t)$ be the position of the prey at time $t$ and $x_B(t)$ denote the position of the current active 
predator. Each of them performs independent Brownian motions with diffusion constants
$D_A$ and $D_B$ respectively, i.e., their positions evolve via
\begin{eqnarray}
\frac{\D x_A(t)}{\D t} &= & \eta_A(t) \label{langeA.1} \\
\frac{\D x_B(t)}{\D t} &= & \eta_B(t) \label{langeB.1}
\end{eqnarray}
where $\eta_A(t)$ and $\eta_B(t)$ are independent Gaussian white noises
with zero mean and the correlators: 
\begin{eqnarray}
\langle \eta_A(t)\eta_A(t')\rangle &=& 2 D_A
\delta(t-t'),\quad  \langle \eta_B(t)\eta_B(t')\rangle= 2 D_B
                                       \delta(t-t'), \label{corr_noise.1} \\ 
  \langle \eta_A(t)\eta_B(t')\rangle &=&0\, .
\label{corr_noise.2}
\end{eqnarray}
The prey has a preferred safe  position, say the origin, and immediately retreats there after surviving an encounter with an active predator.
For simplicity,   we take the initial condition as all particles (predators and prey) located at the origin.
Our goal is to compute the survival probability, $Q(t_0,t)$, of the prey over a time window $[t_0,t]$. 

To make progress, we consider the relative coordinate $x(t)= x_B(t)-x_A(t)$.
Then $x(t)$ also performs  Brownian motion between any two successive
encounters according to
\begin{equation}
\frac{\D x(t)}{\D t}= \eta(t) \equiv \eta_A(t)-\eta_B(t)\, ,
\label{rel_lange.1}
\end{equation}
where the relative noise $\eta(t)$ is again a Gaussian white noise with 
zero mean and its correlator, using Eq. (\ref{corr_noise.1}), is
given by
\begin{equation}
\langle \eta(t)\eta(t')\rangle= 2 (D_A+D_B)\delta(t-t')\, .
\label{rel_corr.1}
\end{equation} 
When the prey encounters the active predator (say at time $t_i$), 
the relative coordinate $x(t_i)$ reaches $0$.  The `relative' particle is then
absorbed  with probability $(1-p)$, corresponding to the situation where the active predator wins. With the complementary probability $p$, the position of the prey  is  reset to the origin and a new predator is designated as
active.
Now, as all predators  perform independent Brownian motions
with diffusion constant $D_B$, the position distribution of any of the passive predators within the swarm at time $t$ is simply
$\ds \frac{ {\rm e}^{-x^2/(4 D_B t)}}{\sqrt{4 \pi D_B t}}$.
Therefore,  since $x_A(t_i)$ resets to $0$ after each encounter with probability $p$,  the relative coordinate is reset from $0$ to $x_B(t_i)$
immediately after the encounter 
where $x_B(t_i)$  is a  Gaussian 
distributed random variable:
\begin{equation}
{\rm Prob.}\left(x_B(t_i) \in [z, z+\D z]\right)=\frac{1}{\sqrt{4 \pi D_B t_i}}\,
{\rm e}^{-z^2/{4 D_B t_i}}\D z\,.
\label{xB_dist.1}
\end{equation}
Crucially, the reset of the relative  co-ordinate to $x_B(t_i)$ is independent of the position of the previous active predator;
this makes manifest the renewal property of the process, which we will use in the following.

\subsection{Survival probability of prey}
We first consider what happens between two successive encounters.
Suppose that the $i$-th encounter happens at time $t_i$.
Let $Q_0(t_i,t)$ denote the probability that the predator and the
prey do not encounter each other up to  time $t>t_i$, after their
last encounter at $t_i$. This can be simply computed as
\begin{eqnarray}
  Q_0(t_i,t)&=& \int_{-\infty}^{\infty} 
                \D z\,  
{\rm erf}\left(\frac{|z|}{\sqrt{4(D_A+D_B)(t-t_i)}}\right)\,
\frac{1}{\sqrt{4\pi D_B t_i}}\, \e^{- z^2/{4 D_B t_i}} \label{survi.1} \\[1ex]
&=& \frac{2}{\pi}\, {\tan}^{-1}
\left[\sqrt{ \frac{\gamma}{\frac{t}{t_i}-1}}\right]\, ,
\label{survi.2}
\end{eqnarray}
where we denote by $\gamma$ the ratio
\begin{equation}
\gamma= \frac{D_B}{D_A+D_B}\, .
\label{def_gamma}
\end{equation}
To understand \eref{survi.1}, note that the integral  is the average
over $z$, the position of the relative coordinate just after the last encounter  at $t_i$, which has the Gaussian distribution \eref{xB_dist.1}. The other factor in the integrand
is  the probability that  the  Brownian motion \eref{rel_lange.1}
does not cross the origin within time $t-t_i$. We  use the well known result \cite{Redner01,BMS13} that for Brownian
motion with diffusion  constant $D$, starting at $x_0$, the probability
of not reaching the origin within time $\tau$ is
${\rm erf}\left(|x_0|/\sqrt{4D\tau}\right)$,
where the error function is defined as 
\begin{equation}
  {\rm erf}(z)= \frac{2}{\sqrt{\pi}}\, \int_0^z \D u\,\e^{-u^2}\;.
\end{equation}
  The resulting integral in \eref{survi.1} may be explicitly evaluated to yield \eref{survi.2}.
Finally, one can easily check that when $t_i \to t$, $Q_0(t_i,t) \to 1$.

We now consider the first-passage probability density $F_0(t_i,t)$, i.e., the probability
that the $(i+1)$-th encounter (the first encounter after the reset at time $t_i$) takes place between time $t$ and $t+\D t$.
This is given by 
\begin{equation}
F_0(t_i,t) \D t=- \partial_t Q_0(t_i,t) \D t\, .
\label{fp.1}
\end{equation}
We note that  $Q_0(t_i,t)$ and $F_0(t_i,t)$ depend on both epochs
$t_i$ and $t$ and hence are non-stationary. However, they can 
be made stationary by introducing the Lamperti transformation \cite{BMS13}, i.e.,
defining the change of time variable $t= {\rm e}^{T}$ or equivalently 
\begin{equation}
 T=\ln t\, .
\label{lamperti_t}
\end{equation}
Then, in terms of the Lamperti time, we have
\begin{eqnarray}
  Q_0(t_i,t) &=&  q_0(T-T_i) \\
  F_0(t_i,t)\D t &=&  f_0(T-T_i)\, \D T\, ,
\label{transform.1}
\end{eqnarray}
where $q_0(T)$ and $f_0(T)$, using equations  (\ref{survi.1}) and (\ref{fp.1}), 
are given explicitly as
\begin{eqnarray}
q_0(T)&= & \frac{2}{\pi} {\tan}^{-1}\left[ 
\sqrt{\frac{\gamma}{\e^T-1}}\right]\, \label{q0T.1} \\[1ex]
  f_0(T)&=& - \frac{\D q_0(T)}{\D T}= 
\frac{\sqrt{\gamma}}{\pi}\, \frac{\e^{T}}{(\e^{T}-1+\gamma)\sqrt{\e^T-1}}\, .
\label{f0T.1}
\end{eqnarray}
Again, one can easily check that when $T\to 0$, $q_0(T) \to 1$.

We next consider the full process over a time window $[t_0,t]$. At the beginning of the time  window, $t_0$,   the process has Gaussian distribution \eref{xB_dist.1}  with $t_i$ = $t_0$, and there may be any number of possible encounters within the window. The survival probability, $Q(t_0,t)$, then satisfies the (first) renewal equation
\begin{equation}
Q(t_0,t)= Q_0(t_0,t)+ p\, \int_{t_0}^t \D t' F_0(t_0,t')\, Q(t',t)\, .
\label{renewal.1}
\end{equation}
The first term in \eref{renewal.1},  $Q_0(t_0,t)$, is the survival probability
without any encounters between predator and prey, i.e., it is simply the survival probability for
Brownian motion with diffusion constant $D_a+D_b$ and  an absorbing target  at the origin, averaged over initial Gaussian distribution \eref{xB_dist.1} at time $t_0$.
The second term in \eref{renewal.1}  integrates over the time, $t'$, of the first
encounter, the factor of $p$ being the probability that the  prey survives this encounter.
The integrand thus contains the probability density for  the time $t'$ of the first encounter,
$F_0(t_0,t')$ multiplied by the survival probability, $Q(t',t)$, over the time window $[t',t]$.

In terms of the Lamperti time $T=\ln(t/t_0)$, \eref{renewal.1} simplifies to
\begin{equation}
q(T)= q_0(T) + p\, \int_0^T \D T' f_0(T') q(T-T')\, .
\label{renewal.2}
\end{equation}
Taking the Laplace transform with respect to $T$, i.e.
\begin{equation}
  \tilde{q}(s)=\int_0^{\infty}\D T\, q(T)\, {\rm e}^{-sT}
  \end{equation}
and using the convolution property 
of the renewal equation (\ref{renewal.2}), we obtain
\begin{equation}
{\tilde q}(s)= \frac{{\tilde q_0}(s)}{1-p {\tilde f_0}(s)}\, ,
\label{LT.1}
\end{equation}
where the Laplace transform of $f_0(T)$ is 
\begin{equation}
  \tilde{f_0}(s)=\int_0^{\infty}\D T\, f_0(T)\, {\rm e}^{-sT}\;.
  \label{f0s}
\end{equation}

\begin{figure}[t]
 \centering
\includegraphics[width=0.9\linewidth]{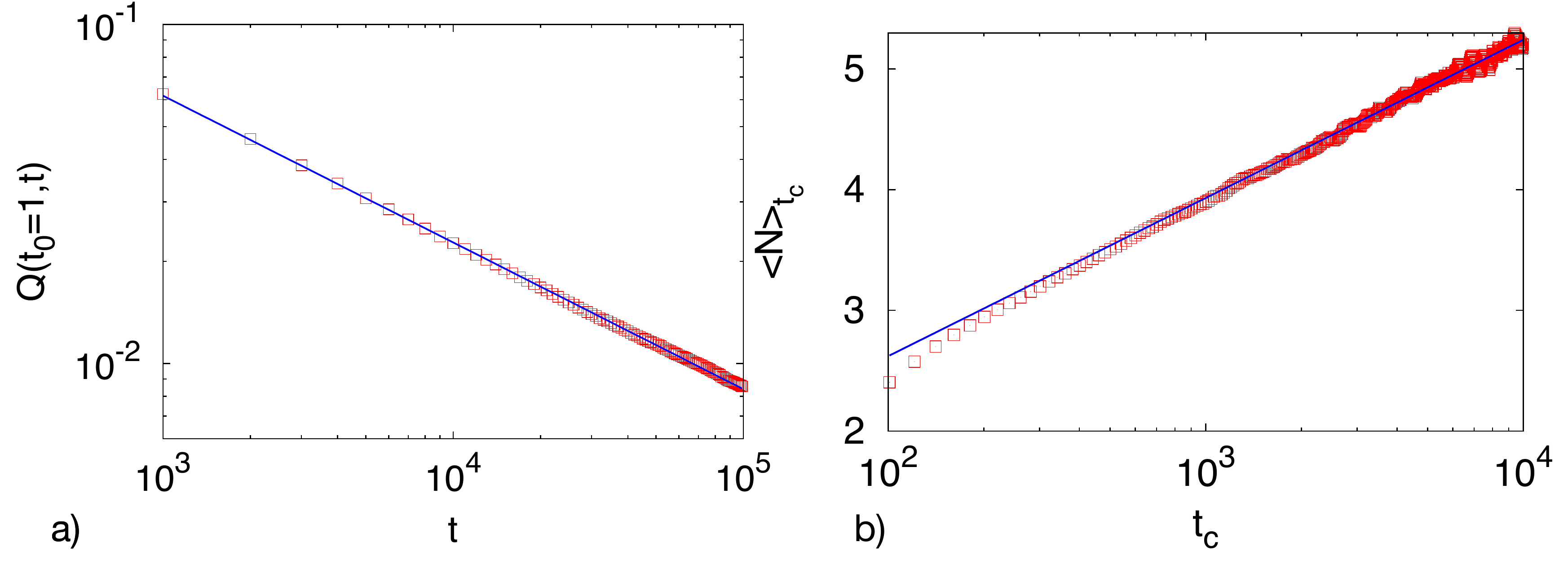}
\caption{{\bf a):} Log-log plot of the survival probability $Q(t_0=1,t)$ vs $t$ for $p=0.25$ and $\gamma=1/2$ (this corresponds to $D_A=D_B$). The symbols correspond to the numerical simulations while the solid blue line corresponds to the fitting function $f(t) = a\, t^{-\theta}$ where $\theta$ is the exact predicted value in Eq. (\ref{theta_exp.1}), where $a$ is just a fitting amplitude. {\bf b):} Log-linear plot of the average number of encounters $\langle N \rangle_{t_c}$ given the capture time $t_c$ vs $t_c$ for $p=0.8$ and $\gamma=1/2$. The symbols correspond to the numerical simulations while the solid blue line corresponds to the fitting function $g(t_c) = A_1 \, \ln t_c$ where $A_1$ is the exact predicted value given in Eq. (\ref{A1A2.1}).}
\label{fig:numerics}
\end{figure}
Using the expression in Eq. (\ref{f0T.1}),  $\tilde{f_0}(s)$ can be explicitly computed as
\begin{equation}
\tilde{f_0}(s)= \frac{\sqrt{\gamma}}{\sqrt{\pi}}\,
\frac{\Gamma(1/2+s)}{\Gamma(1+s)}\,
{}_2F_1\left[1,s+1/2,s+1;1-\gamma\right]
\label{f0s.1}
\end{equation}
where ${}_2F_1\left[a,b,c;z\right]$ is the usual hypergeometric function (See \ref{hypapp}).

The asymptotic large $T$ behaviour of $q(T)$ is determined by the singularity of
$\tilde q(s)$
with largest real part.
If this singularity is a pole at $s=-\theta$ then
\begin{equation}
  q(T)\sim \e^{-\theta T}\;.
\end{equation}
Reverting to real time we then obtain
\begin{equation}
  Q(t_0,t) \sim (t/t_0)^{-\theta}\quad \mbox{for}\quad  t\gg t_0\;.
  \end{equation}
 Thus we identify the exponent  $\theta$ with the dominant pole in  $\tilde{q}(s)$.
 From Eq. (\ref{LT.1}), we  find that the  $\theta$ is given by the smallest positive root
of the transcendental equation, $1-p {\tilde f}_0(-\theta)=0$, i.e.,
\begin{equation}
\frac{1}{p}= \frac{\sqrt{\gamma}}{\sqrt{\pi}}\,
\frac{\Gamma(1/2-\theta)}{\Gamma(1-\theta)}
\,
{}_2F_1\left[1,1/2-\theta,1-\theta; 1-\gamma\right]\;.
\label{theta_exp.1}
\end{equation}
Hence, the exponent $\theta(p,\gamma)$ depends continuously on two parameters, namely,
$0\le p\le 1$ and $0\le \gamma  \le 1$ where $\gamma$ is given by \eref{def_gamma} .
In Figure \ref{fig:numerics} we plot results from numerical simulations,
which show excellent agreement with the analytical predictions.

We now consider some limiting cases where we can develop explicit expressions for the exponent $\theta$. It is easy to check that
for $\gamma=1$, Eq. (\ref{theta_exp.1}) reduces to
\begin{equation}
\frac{1}{p}= \frac{1}{\sqrt{\pi}} \frac{\Gamma(1/2-\theta)}{\Gamma(1-\theta)}\,.
\label{theta_MC.1}
\end{equation}
In fact, in this limit $\gamma \to 1$, it turns out that our model effectively reduces to a model of adaptive persistence studied in a completely
different context in Ref. \cite{MC98} and our expression for the exponent $\theta$ in Eq. (\ref{theta_MC.1}) coincides with that of 
Ref. \cite{MC98} (with $p$ here replaced by $p/2$ in that paper). From the formula (\ref{theta_exp.1})
it is easy to work out the limiting behaviours of the exponent $\theta$
as $p\to 0$ and $p\to 1$. One finds
\begin{eqnarray}
  \label{theta_asymp}  
  \theta(p,\gamma)\to
  \left\{
  \begin{array}{lll}
    & \frac{1}{2} - \frac{\sqrt{\gamma}}{\pi}\, p+ O(p^2)\quad\quad\quad\quad& {\rm as}
      \quad p\to 0 \\[1ex]
    & A(\gamma) (1-p) + O((1-p)^2) \quad & {\rm as}\quad p\to 1\, ,
  \end{array}
      \right.
\end{eqnarray}
where the amplitude $A(\gamma)$ in the second line
in Eq. (\ref{theta_asymp}) can {in principle be computed explicitly 
in terms of hypergeometric function and its derivatives but we do not provide the details here. In Fig. \ref{Fig_theta}, we show a plot of $\theta(p,\gamma)$ as a function of $p$ for $\gamma = 1/2$ obtained from Eq. (\ref{theta_exp.1}), where we also indicate the asymptotic behaviours for $p \to 0$ and $p \to 1$ given in Eq. (\ref{theta_asymp}).

\begin{figure}[t]
\centering
\includegraphics[angle=90,width = 0.6\linewidth]{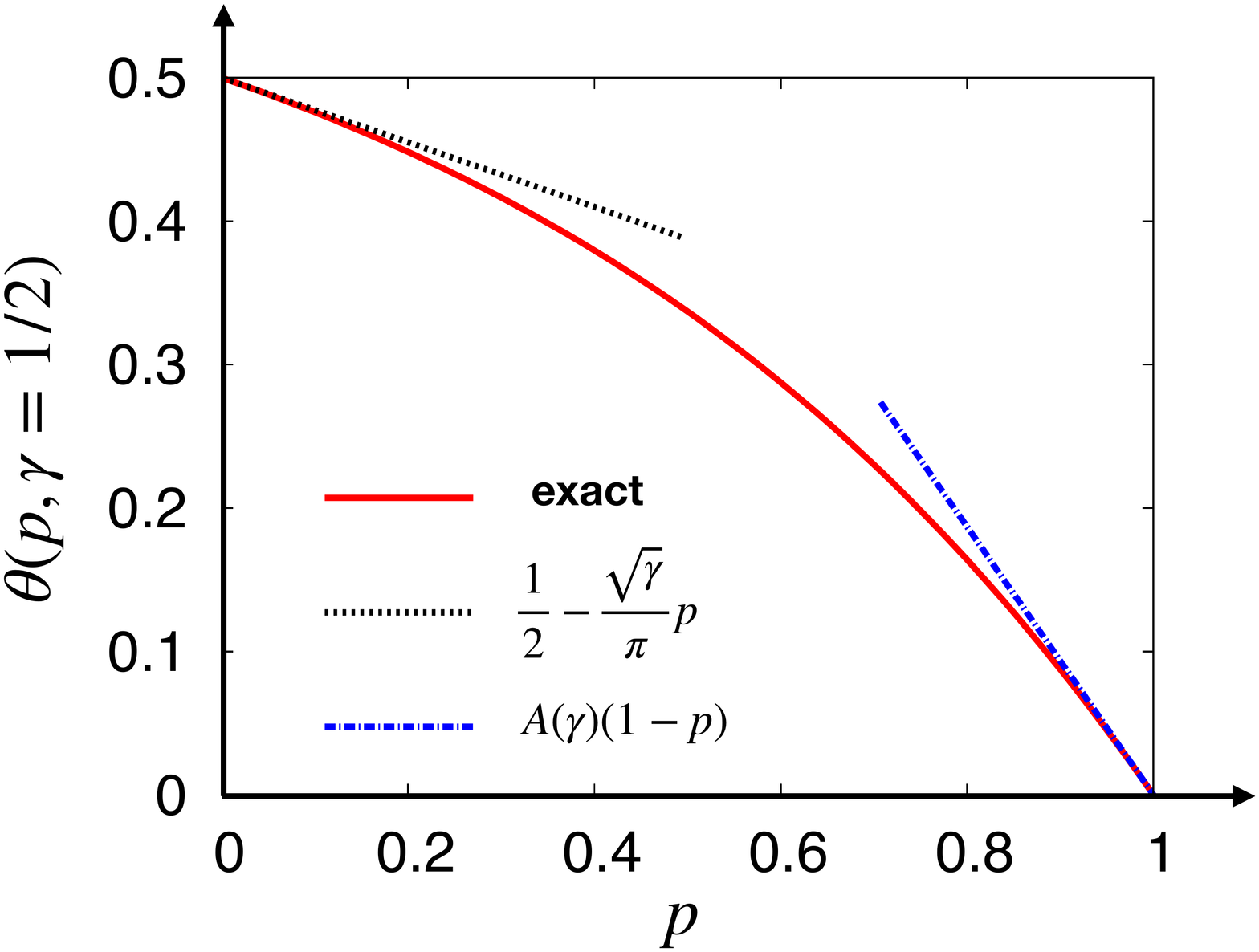}
\caption{Plot of $\theta(p,\gamma)$ vs. $p$ for $\gamma = 1/2$. The solid red line corresponds to the value of $\theta(p,\gamma)$ obtained by inverting numerically the exact relation in (\ref{theta_exp.1}). The dotted line corresponds to the first two terms in the small $p$ expansion of $\theta(p,\gamma=1/2)$ given in the first line of Eq. (\ref{theta_asymp}), while the blue dashed line is the leading behaviour of $\theta(p,\gamma=1/2)$ for $p$ close to $1$ given in the second line of (\ref{theta_asymp}) with $A(\gamma=1/2) = 0.934929\ldots$.} \label{Fig_theta}
\end{figure}

\subsection{Capture time distribution}
The distribution of the capture time $t_c$ of the prey
can also  be simply obtained. In the Lamperti time $T_c=\ln t_c$, this
is 
\begin{equation}
f(T_c)= -q'(T=T_c) \;,
\label{fTc.1}
\end{equation}
where $q(T)$ is given in Eq. (\ref{renewal.2}). Hence its Laplace transform is given by
\begin{equation}
{\tilde f}(s)= \int_0^{\infty} \D T_c\, f(T_c)\, \e^{-s T_c}=1- s {\tilde q}(s)
\label{fs1.1}
\end{equation}
where ${\tilde q}(s)$ is given in Eq. (\ref{LT.1}). This gives
\begin{equation}
{\tilde f}(s)= 1- s \frac{{\tilde q}_0(s)}{1-p {\tilde f}_0(s)}= 
\frac{(1-p){\tilde f}_0(s)}{1-p{\tilde f}_0(s)}\, ,
\label{fs1.2}
\end{equation}
where we used ${\tilde f}_0(s)=1-s {\tilde q}_0(s)$. Note that $\tilde f_0(s)$ is given explicitly in Eq. (\ref{f0s.1}). One may check that $f(T_c)$ is normalized to unity by substituting $s=0$ in Eq. (\ref{fs1.2}), and
using ${\tilde f}_0(0)=1$.

As the expression in Eq. \eref{fs1.2} also has a pole at $s$ satisfying $1-p {\tilde f}_0(s)$,  it follows that the asymptotic behaviour is 
$f(T_c)\sim \e^{-\theta T_c}$ for large $T_c$ with the same $\theta$ given
in Eq. (\ref{theta_exp.1}). In real time $t_c= \e^{T_c}$,
the capture time distribution ${\rm Prob}(t_c)=F(t_c)$ then decays 
for large $t_c$ as a power law
\begin{equation}
F(t_c)= f(T_c) \frac{\D T_c}{\D t_c} \sim (t_c)^{-\theta-1}\, .
\label{Ftc.1}
\end{equation}

\section{Statistics of the number of encounters}

In this section we address another natural question: how many encounters with predators are needed to catch the prey at a given capture time $t_c$?
We begin by  considering the joint probability of $N$ encounters with capture
at the final encounter. The renewal property of the process again facilitates the computation.

\subsection{Joint probability of  number of encounters and capture time}
Consider a typical trajectory of the process till the capture time $t_c$
of the prey, starting at $t_0$. Let the predator and prey encounter
each other at times $\{t_0, t_1, t_2,\ldots, t_N=t_c\}$, i.e.,
there are a total of $N$ encounters till the final capture time. Note that the number of encounters is precisely the number 
of predators needed to catch the prey at time $t_c$.

As usual, it is convenient to work in Lamperti time $T_i=\ln (t_i/t_0)$.
Then the relative process starts at time $T=0$,  and  encounters occur (when $x$ reaches $0$)
at epochs $\{T_1,T_2,\ldots, T_N=T_c\}$. Let us also denote the
intervals between encounters as
\begin{equation}
  \tau_i=T_i-T_{i-1}
\end{equation}
with $T_0=0$ and
\begin{equation}
  T_c= \sum_{i=1}^N \tau_i\;.
  \end{equation}
Thus a `configuration'   is specified by the vector $\vec \tau=\{\tau_1,\tau_2,\ldots, \tau_N\}$, their number $N$ and the final capture time $T_c$, which  are all random variables. The probability of a configuration is given by $P(\vec \tau, N, T_c)$,  the joint distribution of $\vec \tau, N,$ and $T_c$.
This joint distribution can be explicitly written as
\begin{eqnarray}
\lefteqn{  P(\vec \tau, N, T_c)=} \nonumber\\
  && \qquad (1-p)\, p^{N-1}\, f_0(\tau_1)\,f_0(\tau_2)\ldots 
f_0(\tau_N)\,
\delta\left(\tau_1+\tau_2+\ldots +\tau_N-T_c\right)\, ,
\label{pjoint.1}
\end{eqnarray}
where $f_0(\tau)$,  given in Eq. (\ref{f0T.1}),  denotes
the distribution of the time interval between two successive encounters. 
Due to the renewal property, the successive intervals between encounters
are statistically independent, except that they must add up to $T_c$, providing
a global constraint enforced by the delta function in Eq. (\ref{pjoint.1}).
The factor $(1-p)\, p^{N-1}$ in Eq. (\ref{pjoint.1}) reflects the
fact that each of the first $(N-1)$ encounters, where the prey survives, happens with
probability $p$ and the final one, where capture occurs, happens
with probability $(1-p)$.

By integrating over $\tau_i$'s in
Eq. (\ref{pjoint.1}), one obtains the  
joint distribution $P(N,T_c)$ of the number of encounters $N$ 
and the capture time $T_c$
\begin{eqnarray}
\hspace*{-0.7cm}  P(N,T_c)= \frac{(1-p)}{p}\, \int_0^{\infty} %
\prod_{i=1}^N d\tau_i  \left[\prod_{i=1}^N p\, f_0(\tau_i)\right]
\, \delta\left(\tau_1+\tau_2+\ldots +\tau_N-T_c\right)\,.
\label{PNT.1}
\end{eqnarray}
Taking the Laplace transform with respect to $T_c$, one obtains
\begin{eqnarray}
  \tilde P(N,s) &=& \int_0^{\infty} dT_c\, P(N, T_c)\, \e^{-s T_c} = \frac{(1-p)}{p}\, \left[p {\tilde f}_0(s)\right]^N  \, ,
\label{PNT_lap.1}
\end{eqnarray}
where ${\tilde f}_0(s)$ is given by \eref{f0s.1}. Equation \eref{PNT_lap.1} is the main result of this subsection.

One can check that \eref{PNT_lap.1} can be used to 
recover results of Section 2.  Summing $P(N, T_c)$ over $N=1,2,\ldots$, yields
the marginal distribution of the capture
time
\begin{equation}
  f(t_c)=\sum_{N=1}^{\infty} P(N,T_c)\;.
\end{equation}
The Laplace transform of this distribution is obtained by summing Eq. (\ref{PNT_lap.1})
over $N$:
\begin{equation}
  {\tilde f}(s)=
\frac{(1-p){\tilde f}_0(s)}{1-p{\tilde f}_0(s)}\, ,
\label{fs1.3}
\end{equation}
which precisely coincides with Eq. (\ref{fs1.2}). 

\subsection{Moments of $N$ conditioned on the capture time}
\label{sec:mvN}
From the result \eref{PNT_lap.1} it is easy to  compute the conditional distribution of the number of encounters
$P(N|T_c)$, given
the capture time $T_c$. This can be obtained via
\begin{equation}
P(N|T_c)= \frac{P(N,T_c)}{f(T_c)}\, , 
\label{PNC.1}
\end{equation} 
where the Laplace transforms of $P(N,T_c)$ and $f(T_c)$
are given respectively in Eqs. (\ref{PNT_lap.1}) and (\ref{fs1.3}).
Formally inverting these Laplace transforms separately, we can then
write
\begin{equation}
P(N|T_c)= \frac{ {\cal L}^{-1}_{s\to T_c} 
\left[ \left(p\, {\tilde f}_0(s)\right)^N\right]}{p\, {\cal L}^{-1}_{s\to T_c}
\left[\frac{ {\tilde f}_0(s)}{1-p {\tilde f}_0(s)}\right] }\, .
\label{PNC.main}
\end{equation}
where the notation ${\cal L}^{-1}_{s\to T_c}$ means the
Bromwich integral in the complex $s$ plane
\begin{equation}
{\cal L}^{-1}_{s\to T_c}\left[ {\tilde U}(s)\right]
= \int_{\Gamma} \frac{ds}{2\pi i}\,
\e^{s T_c}\, {\tilde U}(s)\, .
\label{brom.1}
\end{equation}


The mean number of encounters 
$\langle N\rangle_{T_c}$, given the capture time $T_c$, can be computed from
\begin{equation}
\langle N\rangle_{T_c}= \sum_{N=1}^{\infty}N\, P(N|T_c)
= \frac{ 
{\cal L}^{-1}_{s\to T_c}
\left[ \frac{ {\tilde f}_0(s)}{(1-p {\tilde f}_0(s))^2}\right]} 
{{\cal L}^{-1}_{s\to T_c}
\left[\frac{ {\tilde f}_0(s)}{1-p {\tilde f}_0(s)}\right] }\, ,
\label{mean.1}
\end{equation}
where we used Eq. (\ref{PNC.main}) and the identity
$\sum_{N=1}^{\infty} N x^N= x/(1-x)^2$. Similarly, the
second moment can be obtained as
\begin{equation}
\langle N^2\rangle_{T_c}= \sum_{N=1}^{\infty} N^2 P(N|T_c) =
\frac{
{\cal L}^{-1}_{s\to T_c}
\left[ \frac{ {\tilde f}_0(s)(1+
p {\tilde f}_0(s))}{(1-p {\tilde f}_0(s))^3}\right]}
{{\cal L}^{-1}_{s\to T_c}
\left[\frac{ {\tilde f}_0(s)}{1-p {\tilde f}_0(s)}\right] }\, .
\label{mom2.1}
\end{equation}

Note that all the Bromwich integrals appearing in
Eqs. (\ref{mean.1}) and (\ref{mom2.1}) are dominated exponentially,
for large $T_c$, by the contribution from the pole at $s=s^*=-\theta$,
where $1-p{\tilde f}_0(s^*)=0$. Thus we need to just calculate the
residues at this pole of all the Bromwich integrals. This can be done
explicitly. Skipping details, we get
\begin{equation}
\langle N\rangle_{T_c}= A_1\, T_c + A_0 + o(1)
\label{mean.2}
\end{equation}
with the prefactors
\begin{equation}
A_1= -\frac{1}{p a_1}=-\frac{1}{p {\tilde f}_0'(s^*)}\,, \quad\,\, 
A_0= \frac{2 a_2}{p a_1^2}-1\, ,
\label{A1A2.1}
\end{equation}
where 
\begin{eqnarray}\label{eq:an}
a_n= \frac{1}{n!}\, {\tilde f}_0^{(n)}(s^*)
\end{eqnarray}
is the $n$-th derivative of $\tilde f_0(s)$, given explicitly in Eq. (\ref{f0s.1}),
evaluated at  the pole $s=s^*$ and divided by $n!$. Note that $a_1= {\tilde f}_0'(s^*)<0$.
Similarly, we get
\begin{equation}
\langle N^2\rangle_{T_c}= B_2\, T_c^2 + B_1\, T_c + B_0 +o(1)\, ,
\label{mom2.2}
\end{equation}
where
\begin{eqnarray}
  B_2&=& \frac{1}{p^2\,a_1^2}\,, \\
  B_1&=& \frac{3}{p^2\,a_1^2}
\left(p\,a_1- 
\frac{2\, a_2}{a_1}\right)\, ,  \\
B_0&=& \frac{1}{p^2\,a_1^2}\left[p^2\, a_1^2-6 p\, a_2 +
6 \left(\frac{2 a_2^2}{a_1^2}
-\frac{a_3}{a_1}\right)\right]\, .
\label{Bs.1}
\end{eqnarray} 
Thus the variance is given by
\begin{equation}
\sigma_N^2= \langle N^2\rangle_{T_c}-\langle N\rangle_{T_c}^2
= (B_2-A_1^2) T_c^2 + (B_1-2 A_1 A_0) T_c + (B_0-A_0^2) +o(1)\, ,
\label{var.1}
\end{equation}
where $A_i$'s and $B_i$'s are given above. Since, $B_2=A_1^2$,
the leading order term $O(T_c^2)$ vanishes in the variance and one gets
for large $T_c$
\begin{equation}
\sigma_N^2 = B\, T_c + O(1)\, ,
\label{var.2}
\end{equation}
where the prefactor $B$ is given explicitly by
\begin{equation}
B= B_1-2 A_1\, A_0=\frac{1}{p a_1}- \frac{2a_2}{p^2 a_1^3}
= \frac{1}{p {\tilde f}_0'(s^*)}- \frac{{\tilde f}_0''(s^*)}{p^2 
({\tilde f}_0'(s^*))^3}\, ,
\label{var.amp}
\end{equation}
with ${\tilde f_0}(s)$ given explicitly in Eq. (\ref{f0s.1})
and $s^*$ is determined from ${\tilde f}_0(s^*)=1/p$.

For general $(p,\gamma)$ the amplitude $B$ in Eq. (\ref{var.amp})
has a complicated expression, in particular since there is no
explicit expression for $s^*$. For $\gamma=1$,  ${\tilde f}_0(s)$ simplifies to
\begin{equation}
{\tilde f}_0(s)= \frac{1}{\sqrt{\pi}}\, \frac{\Gamma(s+1/2)}{\Gamma(s+1)}\, .
\label{g1.1}
\end{equation}
In this case, the expression for $B$ in Eq. (\ref{var.amp}) can be 
explicitly evaluated as
\begin{equation}
B= \frac{\psi ^{(1)}(s^*+1)-\psi
  ^{(1)}\left(s^*+\frac{1}{2}\right)}{(\psi ^{(0)}(s^*+1)-\psi ^{(0)}\left(s^*+\frac{1}{2}\right))^3}\,,
\label{Bg1.1}
\end{equation}
where $\psi^{(m)}(z) $ is the polygamma function of index $m$, i.e., the $(m+1)$-th derivative of $\ln \Gamma(z)$ and $s^*$ is again determined from ${\tilde f}_0(s^*)=1/p$
with ${\tilde f}_0(s)$ given in Eq. (\ref{g1.1}).

Note that in terms of the original time $t_c= \e^{T_c}$,
the results for the mean and the variance of $N$, respectively
in equations. (\ref{mean.2}) and (\ref{var.2}) translate, for large
$t_c$, to
\begin{equation}
\langle N\rangle_{t_c}= A_1 \ln t_c +O(1)\, , \quad {\rm and}\quad 
\sigma^2_N= B\, \ln t_c + O(1)\, ,
\label{mean_var.1}
\end{equation}
indicating a rather slow logarithmic growth in the original capture time
$t_c$.  
Note that the mean number of resettings grows logarithmically with time, as opposed to linearly in standard
resetting processes where resetting occurs at a constant rate. The reason behind this slow growth can be traced back to the
fact that the interval between successive resettings/encounters is power-law distributed in this problem.


\subsection{The large deviation form of the conditional
distribution $P(N|T_c)$}

We now consider
the full conditional distribution $P(N|T_c)$ in equation (\ref{PNC.main}) in
the limit when $N$ is large, $T_c$ is large, but with the ratio
$z= N/T_c$ fixed. We will derive a large deviation form for $P(N|T_c)$.

The denominator in equation (\ref{PNC.main})
can again be evaluated for large $T_c$ by computing the
residue at the pole $s=s^*$ and one gets, to leading order for large $T_c$
\begin{equation}
p {\cal L}^{-1}_{s\to T_c}
\left[\frac{ {\tilde f}_0(s)}{1-p {\tilde f}_0(s)}\right]
\approx \frac{1}{-p \, {\tilde f}_0'(s^*)}\, \e^{s^* T_c}\, ,
\label{denom.1}
\end{equation}
where $s^*$ is given by ${\tilde f}_0(s^*)=1/p$. The numerator
in equation (\ref{PNC.main}) can be evaluated, for large $N$, by the
saddle point method. Ignoring pre-exponential terms, this gives
\begin{equation}
{\cal L}^{-1}_{s\to T_c}
\left[ \left(p\, {\tilde f}_0(s)\right)^N\right]\sim
\exp\left[T_c\, {\min}_s[ s + z \ln (p {\tilde f}_0(s)]\right]\, ,
\label{saddle.1}
\end{equation}
where the minimum occurs at the saddle point with real $s= s_0$ and
$z=N/T_c$. Taking the ratio of the numerator and denominator in
equation (\ref{PNC.main}), we then get for fixed $z$, the following
large deviation behaviour in the limit $N\to \infty$, $T_c\to \infty$
but with $z=N/T_c$ fixed
\begin{equation}
P(N|T_c)\sim \e^{- T_c\, \Phi\left(z=\frac{N}{T_c}\right)}\, ,
\label{ldv.1}
\end{equation}
where the rate function $\Phi(z)$ is given by
\begin{equation}
\Phi(z)= -{\min}_s\left[ s + z \ln (p\, {\tilde f}_0(s))\right]+s^* = -{\min}_s\left[ s + z \ln (p\, {\tilde f}_0(s))\right] - \theta\,,
\label{ldv.2}
\end{equation}
where we have used $s^* = -\theta$, which is the solution of $1 - p \tilde f_0(s^*) = 0$.

While it is difficult to compute this rate function $\Phi(z)$ exactly for all $z$, one can derive its asymptotic behaviours
as $z \to 0$ and $z \to \infty$ and also its behaviour near its minimum. Suppose that the minimum over $s$ in equation (\ref{ldv.2}) occurs
at $s=s_0$. Defining $W(s) = s + z \ln (p\, {\tilde f}_0(s))$, and setting $W'(s_0) = 0$ gives 
\begin{eqnarray}\label{def_s0}
-\frac{{\tilde f}'_0(s)}{{\tilde f}_0(s)} \Bigg \vert_{s=s_0} = \frac{1}{z} \;,
\end{eqnarray}
which determines $s_0$ upon using $\tilde f_0(s)$ from equation (\ref{f0s.1}). In Fig. \ref{Fig_s0}, we give a plot of the function $-{{\tilde f}'_0(s)}/{{\tilde f}_0(s)}$ vs $s$ for $\gamma=1$. From this figure \ref{Fig_s0}, it is clear that when $z \to 0$, $s_0 \to -1/2$, while when $z \to \infty$, $s_0 \to \infty$.   
\begin{figure}[t]
 \centering
\includegraphics[width=0.6\linewidth]{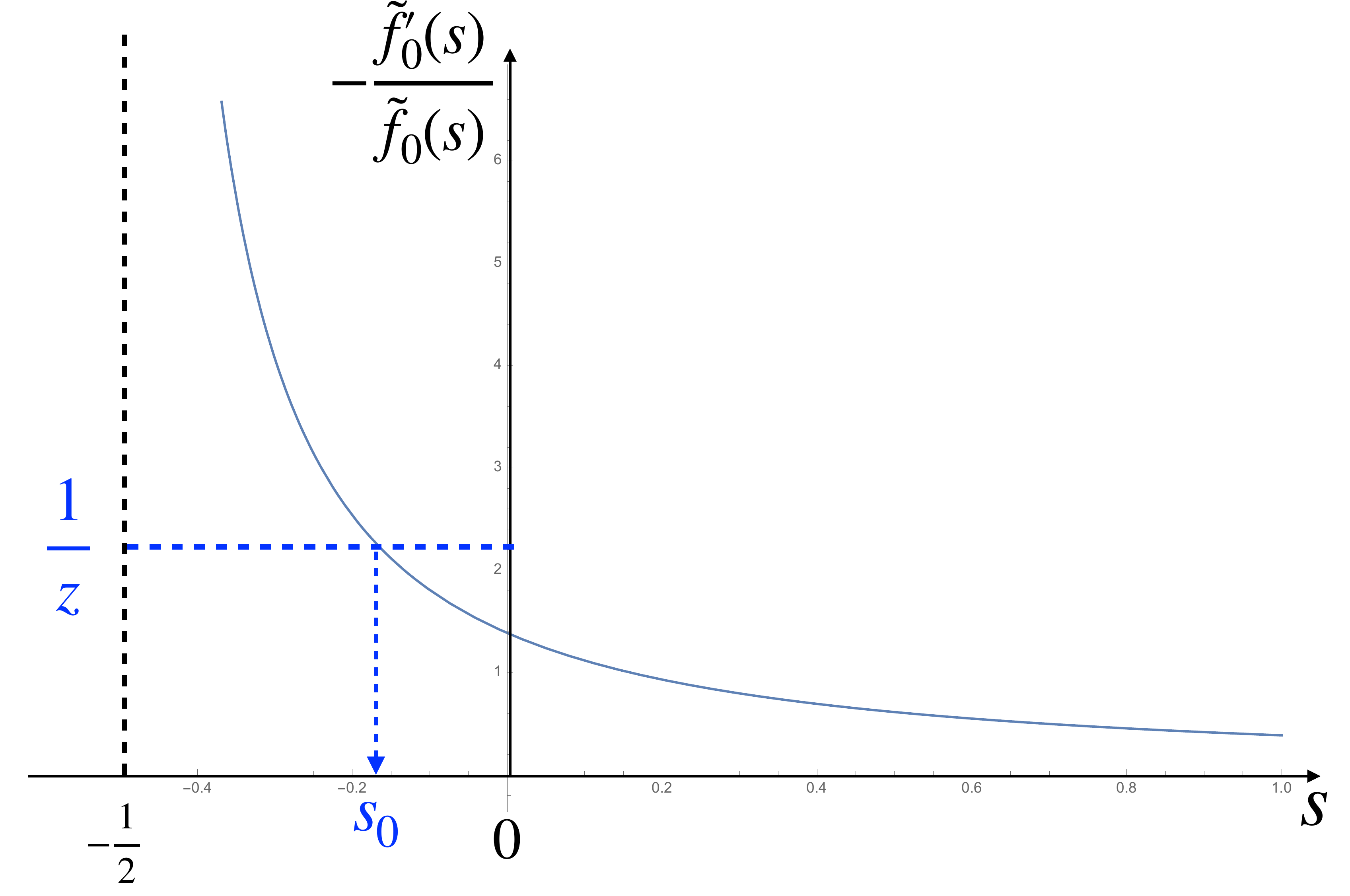}
\caption{Plot of $-{{\tilde f}'_0(s)}/{{\tilde f}_0(s)}$ vs $s$ for $\gamma=1$, using $\tilde f_0(s)$ from equation (\ref{g1.1}). 
The function diverges as $s \to -1/2$ and decreases monotonically for increasing $s$. The location of the saddle point $s_0$ is given by the value of $s$ at which this curve crosses the value $1/z$, as indicated in the figure. For other values of $0<\gamma < 1$ the behaviour is qualitatively similar.}\label{Fig_s0}
\end{figure}

 \subsubsection{The limit $z\to 0$.}
 We first consider the limit $z \to 0$. From Fig. (\ref{Fig_s0}), we see that $s_0\to -1/2$ as $z\to 0$ (consequently $1/z$ diverges) and hence we set $s_0 = -1/2 + \epsilon$. Expanding the function for small $\epsilon$, we get from Eq. (\ref{def_s0})
\begin{eqnarray} \label{z_small}
z = \epsilon + O(\epsilon^2) \;.
\end{eqnarray}
Inverting, for small $z$, yields
\begin{eqnarray} \label{small_eps}
\epsilon = z + O(z^2) \;.
\end{eqnarray}
We also find that, to leading order in $\epsilon$ 
\begin{eqnarray} \label{f0_small}
\tilde f_0\left(s_0 = -\frac{1}{2} + \epsilon\right) = \frac{\sqrt{\gamma}}{\pi \epsilon} + O(1) \;.
\end{eqnarray}
We substitute $s_0 = -1/2 + \epsilon$, $\epsilon = z + O(z^2)$ and $\tilde f_0(s_0=-1/2+\epsilon)$ from Eq. (\ref{f0_small}) in the expression for $\Phi(z)$ in Eq. (\ref{ldv.2}). This gives, as $z \to 0$, 
\begin{eqnarray} \label{phi_small}
\Phi(z) = \frac{1}{2}-\theta + z \ln z - z\,\ln \left( \frac{p\,e\,\sqrt{\gamma}}{\pi}\right) + O(z^2) \;.
\end{eqnarray}
Thus, as $z \to 0$, the rate function approaches a constant $\Phi(z) \to 1/2 -\theta > 0$. This can be understood by setting, for example, $N=1$ in Eq. (\ref{PNC.main}). For $N=1$, the numerator behaves as $\sim \e^{-T/2}$ since $f_0(T) \sim \e^{-T/2}$ from Eq. (\ref{f0T.1}). The denominator in Eq. (\ref{PNC.main}) scales, for large $T$, as $\sim \e^{-\theta T}$. Hence the ratio, for $N=1$, scales as $\sim \e^{-(1/2-\theta)\,T}$, indicating, from the large deviation form in Eq. (\ref{ldv.1}), that $\Phi(0) = 1/2 - \theta$. Note that this leading term does depend on both $p$ and $\gamma$. However, the first sub-leading term $z \ln z$ is universal, i.e., independent of $p$ and $\gamma$.

\subsubsection{The limit $z\to \infty$.}
We now turn to the opposite $z \to \infty$ limit. From Fig. \ref{Fig_s0}, it is clear that $s_0 \to \infty$ for large $z$. Expanding $\tilde f_0(s)$ in Eq. (\ref{f0s.1}) for large $s$, we find 
\begin{eqnarray}\label{f0_large}
\tilde f_0(s) = \frac{1}{\sqrt{\pi \gamma \,s}} + \frac{3 \gamma-4}{8 \sqrt{\pi} \gamma^{3/2} s^{3/2}} + O(1/s^{5/2}) \;.
\end{eqnarray}
Substituting this asymptotic behaviour in Eq. (\ref{def_s0}) gives, for large $z$
\begin{eqnarray} \label{z_large}
z= 2 s_0 + \frac{4-3\gamma}{2\gamma} + O(1/s_0) \;.
\end{eqnarray}
Inverting this relation, we get
\begin{eqnarray}\label{s_large}
s_0 = \frac{z}{2} - \frac{4-3\gamma}{4 \gamma} + O(1/z) \;.
\end{eqnarray}
Substituting these behaviours in the expression for the rate function $\Phi(z)$ in (\ref{ldv.2}), gives, for large $z$
\begin{eqnarray} \label{phi_large}
\Phi(z) = \frac{z}{2} \ln z - z \ln\left( p \sqrt{\frac{2\,e}{\gamma \pi}}\right) + \frac{4-3\gamma}{4\gamma} - \theta + O(1/z) \;.
\end{eqnarray}
Note that the leading term is universal, i.e., independent of $p$ and $\gamma$.

\subsubsection{The behaviour close to the minimum of $\Phi(z)$.}
From the expression of $\Phi(z)$ in Eq. (\ref{ldv.2}), it is clear that when $s_0 \to s^*$ where $f_0(s^*) = 1/p$, the rate function approaches $-s^* - \theta = 0$, which is the minimum value of $\Phi(z)$, since the rate function is necessarily non-negative. We therefore set $s_0 = s^* + \epsilon$, with $|\epsilon| \ll 1$ in order
to study the behaviour of $\Phi(z)$ in the vicinity of its minimum. Expanding the saddle point equation (\ref{def_s0}) for small $|\epsilon|$ and using $p \tilde f_0(s^*) = 1$, we get
\begin{eqnarray} \label{z_mini}
\frac{1}{z} = - p\tilde f_0'(s^*)+p^2 \left((\tilde f_0'(s^*))^2 - \frac{1}{p} \tilde f_0''(s^*)\right)\epsilon + O(\epsilon^2) \;.
\end{eqnarray} 
This indicates that, exactly at $\epsilon = 0$, $z \to z^* = -1/(p \tilde f_0'(s^*)) = A_1$ where $A_1$ has already been defined in Eq. (\ref{A1A2.1}). Setting $z = A_1 + \delta$ and inverting the relation (\ref{z_mini}) we obtain
\begin{eqnarray} \label{eps_mini}
\epsilon = - \frac{1}{C\,A_1^2} \delta + O(\delta^2) \quad, \quad {\rm where} \quad C = (p\,\tilde f_0'(s^*))^2 - p \tilde f_0''(s^*) \;.
\end{eqnarray} 
We now expand $\Phi(z)$ in Eq. (\ref{ldv.2}) around $z=A_1$ and express it as a function of $\delta$ only. To leading order we get
\begin{eqnarray} \label{Phiz_quad}
\Phi(z) = \frac{1}{2C^2A_1^5} \left(1 - 2 C A_1^2 - p A_1^2 \tilde f_0''(s) \right)\, \delta^2 + O(\delta^3) \;.
\end{eqnarray}
Substituting the expression for $A_1$ and $C$ and simplifying, we find a leading quadratic behaviour of $\Phi(z)$ around $z = A_1$
\begin{equation}
\Phi(z)\approx \frac{1}{2B}(z-A_1)^2\, \quad {\rm as} \,\, z\to A_1\, ,
\label{ldv_quad.1}
\end{equation}
where $A_1$ and $B$ are respectively in Eqs. (\ref{A1A2.1})
and (\ref{var.amp}). Note that this quadratic behaviour of the rate function indicates that the typical fluctuations of $N$, for fixed $T_c$,
are Gaussian distributed with mean $\langle N\rangle_{T_c}\approx A_1 T_c$
and variance $\sigma_N^2 \approx B\, T_c$, as we had already derived in Section \ref{sec:mvN},
Thus the mean and the variance from the expansion of the 
large deviation function near its minimum coincide perfectly with their expressions obtained directly in Eqs. (\ref{mean.2}) and (\ref{var.2}) respectively. 

The asymptotic behaviours of $\Phi(z)$ can be summarised as follows
\begin{eqnarray}\label{summary}
\Phi(z) = \left\{
\begin{array}{lll}
 \frac{1}{2}-\theta + z \ln z - z\,\ln \left( \frac{p\,e\,\sqrt{\gamma}}{\pi}\right) + O(z^2) \quad & z \to 0 \\[1ex]
  \frac{1}{2B}(z-A_1)^2\, \quad &z\to A_1 \\[1ex]
  \frac{z}{2} \ln z - z \ln\left( p \sqrt{\frac{2\,e}{\gamma \pi}}\right) + \frac{4-3\gamma}{4\gamma} - \theta + O(1/z)  \quad & z \to \infty \;,
\end{array}
  \right.
\end{eqnarray}
where $A_1$ and $B$ are respectively in Eqs. (\ref{A1A2.1}) and (\ref{var.amp}). 
A plot of this function is provided in Fig. (\ref{Fig_large_dev}) for $\gamma=1$ and $p=1/2$, 
for which $\theta = 0.300568 \ldots\;$. In this case, from Eqs. (\ref{A1A2.1}) and (\ref{var.amp}),
we get $A_1= 0.244956\ldots$ and $B=0.346483\ldots\;$.
In this plot, we also provide the asymptotic behaviours for $z \to 0$, $z \to \infty$, 
as well as the quadratic behaviour near the minimum $z=A_1=0.244956\ldots$ in Eq. (\ref{summary}),
finding excellent agreement. 

\begin{figure}[t]
  \centering
\includegraphics[angle=90,width = 0.7\linewidth]{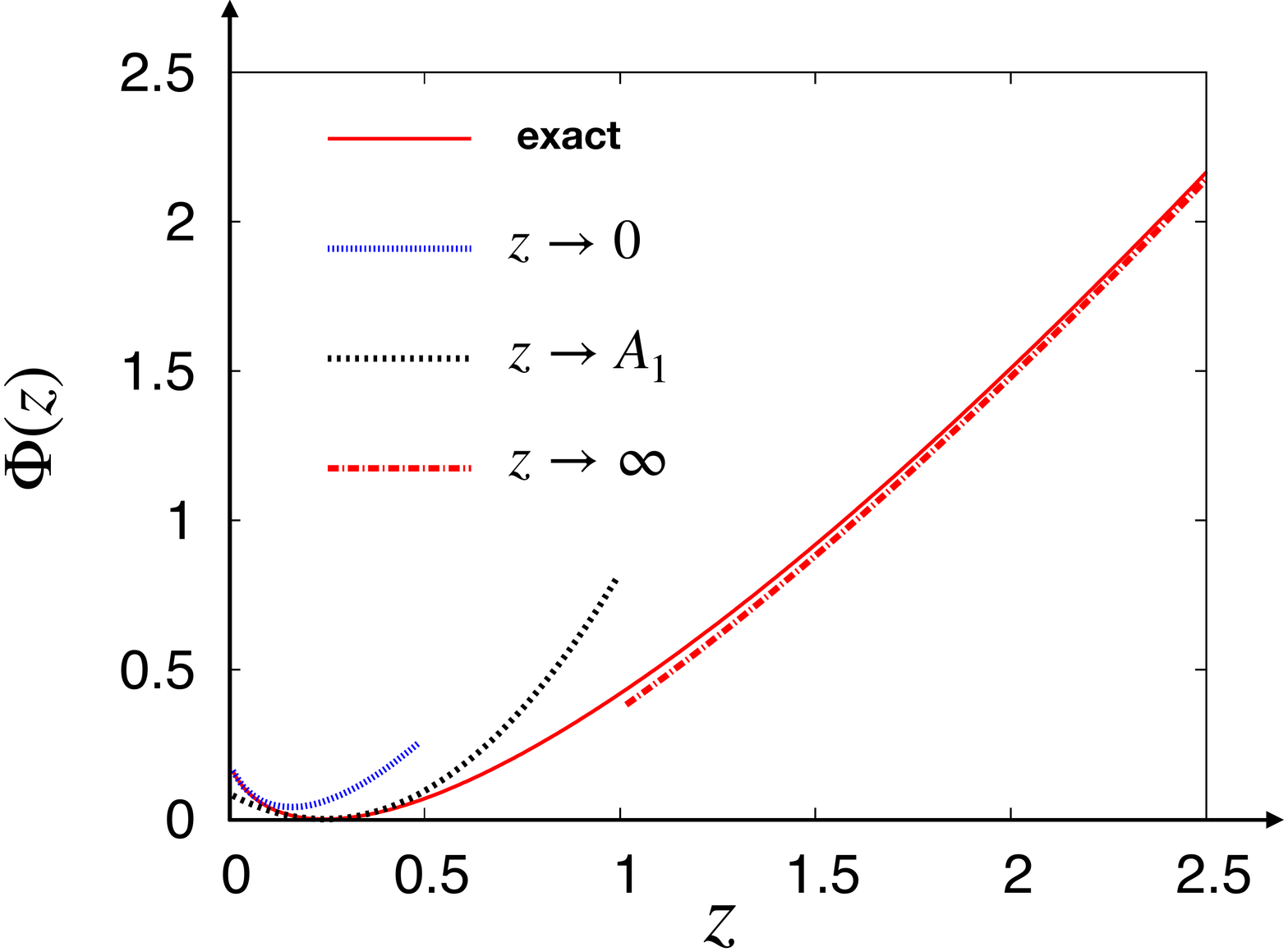}
\caption{The rate function $\Phi(z)$ vs $z$ for $\gamma=1$ and $p=1/2$ for which 
$\theta = 0.300568 \ldots\;$. 
The asymptotic behaviours of $\Phi(z)$ for $z \to 0$, $z \to \infty$,
as well as the quadratic behaviour in Eq. (\ref{summary}), namely $\Phi(z)\approx \frac{1}{2B}(z-A_1)^2 $ near $z=A_1=0.244956\ldots$
(and
with $B=0.346483\ldots$), are also plotted.}
\label{Fig_large_dev}
\end{figure}

Finally, reverting back to the original time $t_c=\e^{T_c}$, we see
that the large deviation form in Eq. (\ref{ldv.1}) translates into
an anomalous form in $t_c$, namely,
\begin{equation}
P(N|t_c)\sim t_c^{- \Phi\left(\frac{N}{\ln t_c}=z\right)}\, .
\label{ldv_tc.1}
\end{equation}
Thus $\Phi(z)$ acts like an `effective' exponent if one observes $P(N|t_c)$ as a function of $t_c$ for large $t_c$ and fixed $N$. This type of ``anomalous large deviation behaviour'' was also found in other contexts such as in the study of the persistence/survival probability of a stationary Gaussian process ~\cite{MB98,BMS13,SM07,SM08,PS18}. 

\section{Conclusion}
In this work we have considered a simple predator-prey model
where on encounters with an active predator, the prey has a probability of surviving and escaping back to the origin.
The model exhibits a
survival probability  decaying as a power law with a nontrivial exponent
$\theta(p,\gamma)$. This exponent  depends continuously on the escape probability $p$ and the parameter $\gamma$ \eref{def_gamma} which is a ratio involving the diffusion constants of predator and prey. Moreover, we have studied the distribution of the number of encounters $N$ conditioned  on capture at time $t_c$.

The presence of a swarm of predators from which a new active predator is selected after
each encounter implies a renewal form for the survival probability \eref{renewal.1}.
This renewal property facilitates  computation of the  survival probability
and other quantities we have considered.

It would be natural to consider a single predator and prey, with the same escape probability $p$ after an 
encounter but with the relative separation then being reset to the position of the prey. However, in that case 
the corresponding equation for the survival probability would contain a memory of the position of the predator 
at the encounter, and such memory dependence poses an open problem. In this case, we expect again that the 
exponent $\theta$ will depend continuously on $p$ and $\gamma$, though its expression will be different from 
the result in Eq. (\ref{theta_exp.1}). Our preliminary numerical simulations confirm this conclusion.
Another natural modification of the model would be to consider multiple predators such that once the
prey survives an encounter and returns to its nest, the status `active' is accorded to
the predator who is currently closest to the prey at the origin following its return. 
It would be interesting to explore these cases and other generalisations of the model.

Stripping away the context predator and prey, the process reduces to a positive diffusive field \eref{rel_lange.1}, i.e.,  
the separation, which on reaching the origin is reset with probability $p$ to a value drawn from the time-dependent 
distribution \eref{xB_dist.1}. In this light, the connection with the model of \cite{MC98} and various problems of 
persistence of a fluctuating field \cite{MB98,BMS13,SM07,SM08,PS18} becomes apparent.

The origin of the slow decay of the survival probability is that at each resetting event the separation of predator 
and prey is reset to a time-dependent distribution \eref{xB_dist.1} which is a Gaussian whose width increases with 
time as $\sqrt{t}$. Thus at each reset the typical separation increases. It would be of interest to see if 
other choices of the 
resetting distribution of the separation could lead to nontrival forms for the survival probability, e.g., a stretched 
exponential. Another interesting extension would be to consider `interacting' predators with
a short-range interaction. If the positions of such interacting walkers typically scale as
$\sqrt{t}$, we believe that much of the behaviour found here for noninteracting predators would still hold, e.g., the
exponent characterizing the decay of the survival probability will still be a continuous function
of the parameters of the model. Similarly, the large deviation form for the number of encounters in Eq. (\ref{ldv_tc.0}), given
the capture time $t_c$, is also expected to hold albeit with a different rate function $\Phi(z)$.
An example of such short-ranged interacting predators where the positions still scale diffusively as $\sqrt{t}$
is given by the nonintersecting Brownian motions (also known as vicious walkers).
It would be interesting to compute the survival probability
exponent $\theta$ or the rate function $\Phi(z)$ analytically for this example.

\ack
MRE would like to thank LPTMS for a Visiting Professorship.

\appendix
\section{Derivation of formula \eref{f0s.1}} \label{hypapp}
Here we compute  the Laplace transform of the first passage time distribution
\begin{equation}
  \tilde{f_0}(s)=\int_0^{\infty}\D T\, f_0(T)\, \e^{-sT}\;.
  \label{f0s}
\end{equation}

Using expression Eq. (\ref{f0T.1}) for $f_0(T)$,  we obtain
\begin{equation}
  \tilde{f_0}(s)= \frac{ \gamma^{1/2}}{\pi} \int_0^{\infty}\D T\,\frac{ {\rm e}^T}{({\rm e}^T -1)^{1/2}}
  \frac{ {\rm e}^{-sT}}{({\rm e}^T -1+\gamma)}\;.
  \label{f0s2}
\end{equation}
Changing  integration variable to $x = {\rm e}^{-T}$ yields
\begin{equation}
  \tilde{f_0}(s)= \frac{ \gamma^{1/2}}{\pi} \int_0^{1}\D x\,\frac{ x^{s-1/2}}{(1-x)^{1/2}(1-(1-\gamma)x)}\, .
  \label{f0s3}
\end{equation}
We now compare  \eref{f0s3}  with the integral representation of the hypergeometric function
\begin{equation}
  {\rm B}(b, c-b)\, {}_2F_1\left[a,b,c; z\right] = \int_0^1 \D x\, x^{b-1}(1-x)^{c-b-1}(1-zx)^{-a}\;,
\end{equation}
where  the Beta function  is defined, as usual, by
\begin{equation}
  {\rm B}(u,v) = \frac{ \Gamma(u) \Gamma(v)}{\Gamma(u+v)}\;,
\end{equation}
and $\Gamma(u)$ is the usual gamma function.
We identify $a=1$, $b=s+1/2$, $c=s+1$, $z = 1-\gamma$ and hence deduce
\begin{equation}
\tilde{f_0}(s) = \frac{ \gamma^{1/2}}{\pi^{1/2}}
\frac{\Gamma(1/2+s)}{\Gamma(1+s)}\,
{}_2F_1\left[1,s+1/2,s+1;1-\gamma\right]\;.
\end{equation}

\end{document}